\documentclass[aps,showpacs,nofootinbib,preprintnumbers,nobalancelastpage]{revtex4}

\usepackage[english]{babel}
\usepackage{amsmath,amssymb}
\usepackage[dvips]{graphicx}
\usepackage[sort&compress]{natbib}
\usepackage{amsfonts}
\usepackage{color}

\DeclareMathAlphabet{\mathsc}{OT1}{cmr}{m}{sc}

\allowdisplaybreaks
\newcommand{\ie} {{\it i.e.}}

\def\321{SU(3) $\otimes$ SU(2) $\otimes$ U(1)}

\def\lsim{\raise0.3ex\hbox{$\;<$\kern-0.75em\raise-1.1ex\hbox{$\sim\;$}}}
\def\gsim{\raise0.3ex\hbox{$\;>$\kern-0.75em\raise-1.1ex\hbox{$\sim\;$}}}

\newcommand{\flux}[2][]{\ensuremath{\ifthenelse{\equal{#1}{}}{}{^{#1}\!}\mathit{#2}}}

\newcommand{\sla}[1]{/\!\!\!\!#1}

\begin{document}
%\preprint{XXXXXXXX}

\title{Update of the Present Bounds on New Neutral Vector Resonances from Electroweak 
Gauge Boson Pair Production at the LHC}
\author{J.\ Gonzalez--Fraile}
\email{fraile@ecm.ub.es}
\affiliation{%
  Departament d'Estructura i Constituents de la Mat\`eria and
  ICC-UB, Universitat de Barcelona, 647 Diagonal, E-08028 Barcelona,
  Spain}

%---------------------------------------------------------------------
\begin{abstract}

The model independent bounds on new neutral vector resonances masses, couplings and widths
presented at arxiv:1112.0316~\cite{Eboli:2011ye} are updated with an integrated luminosity of $\mathcal{L}=4.7\ \mbox{fb}^{-1}$
from ATLAS and $\mathcal{L}=4.6\ \mbox{fb}^{-1}$ from CMS.
These exclusion limits correspond to the most stringent existing bounds
on the production of new neutral spin--1 resonances that decay to electroweak gauge boson pairs
and that are associated to the electroweak symmetry breaking sector in several extensions of the
Standard Model.

\end{abstract}
%---------------------------------------------------------------------

 % Elementary particle processes

\maketitle

%%%%%%%%%%%%%%%%%%%%%%%%%%%%%%%%%%%%%%%%%%%%%%%%%%%%%%%%%%%%%%%%%%%%%%

\section{Introduction}

In this talk I update the bounds on new neutral vector resonances
($Z^\prime$) associated to the EWSB~\cite{Eboli:2011ye}, that are common in many
extensions of the Standard Model (SM).
The updated bounds are derived using data from ATLAS
(with integrated luminosity of $\mathcal{L}=4.7\ \mbox{fb}^{-1}$) and CMS
(with integrated luminosity of $\mathcal{L}=4.6\ \mbox{fb}^{-1}$) on
$W^+W^-$ pair production. Including a $Z^\prime$ the process is
\begin{equation}
  p p \to Z^\prime \to W^+ W^- \to \ell^+ \ell^{\prime -} \, \sla{E}_T
\label{eq:proc}
\end{equation}
where $\ell$ and $\ell^\prime$ stand for electrons and muons.
The bounds are presented in a model independent way as
constraints on the relevant spin-1 boson effective couplings, mass and
width. These exclusion limits correspond to the most stringent direct bounds that we are aware of on the
production of a $Z^\prime$ that decay to electroweak gauge boson pairs.
As an example, a $Z^\prime$ coupling
with SM strength to light quarks and saturating the $W^+W^-$ 
partial wave amplitudes can be excluded at the 2$\sigma$ level for masses lighter
than $\simeq 2$ TeV. \smallskip

After describing the basic details
of the model independent framework for the $Z^\prime$
properties as well as the details of the analyses in Section~\ref{sec:frame},
I present the model independent results using the updated data sets in
Section~\ref{sec:results}.\smallskip

For the complete details of the simulations and analyses, as well as an extended
discussion of the bounds and an example of how to translate the model independent
bounds to a given model, the reader is referred to the original publication~\cite{Eboli:2011ye}
that this update is based on.

%%%%%%%%%%%%%%%%%%%%%%%%%%%%%%%%%%%%%%%%%%%%%%%%%%%%%%%%%%%%%%%%%%%%%%
\section{Framework and analyses details}
\label{sec:frame}
%%%%%%%%%%%%%%%%%%%%%%%%%%%%%%%%%%%%%%%%%%%%%%%%%%%%%%%%%%%%%%%%%%%%%%

In the analysis of the present bounds on the production of new neutral
vector resonances we work in a framework ~\cite{Alves:2009aa,Eboli:2011bq}
where the relevant coupling of the process (\ref{eq:proc}),
the mass and the width are considered free parameters of the study.
Inspired by models where the interactions of the new $Z^\prime$
are due to its mixing with the SM gauge bosons, we also assume that new vector
resonance coupling to light quarks and $W^+W^-$ pairs have
the same Lorentz structure as the ones of the SM.
\smallskip

Defining the normalization factor ${g_{Z^\prime WW}}_{max}$ as the 
$Z^\prime W^+ W^-$ coupling that saturates the partial wave amplitude
for the process $W^+ W^- \to W^+ W^-$ by the exchange of a $Z^\prime$,
~\cite{Birkedal:2004au}
\begin{equation}
{g_{Z^\prime WW}}_{max}=g_{ZWW}\, \frac{M_Z}{\sqrt{3}M_{Z^\prime}} 
\label{eq:gwwvmax}
\end{equation}
where $g_{ZWW}=g~ c_W$ is the strength of the SM triple gauge boson
coupling, $g$ is the $SU(2)_L$ coupling constant and $c_W$ is the cosine
of the weak mixing angle, we can define the relevant product of couplings of process~(\ref{eq:proc})
as the combination:
\begin{equation}
G=\left(\frac{g_{Z^\prime q\bar q}}{g_{Zq\bar q}} \right)\,
\left(\frac{g_{Z^\prime WW}}{{g_{Z^\prime WW}}_{max}}\right)\, ,
\label{eq:G}
\end{equation}
here $g_{Z'q\bar q}$ and $g_{Z'WW}$ are the coupling constants of $Z^\prime$ to light quarks and $W^+W^-$, respectively,
and $g_{Z q\bar q}=g/c_W$.
\smallskip

In this approach we treat $G$, the $Z^\prime$ width and its mass
as free parameters, but for consistency with the decay of the $Z^\prime$
to light quarks and $W^+W^-$ pairs we get the constraint~\cite{Alves:2009aa}:
\begin{eqnarray}
&&\Gamma_{Z^\prime}\; >\;0.27\, |G| \, 
\,\left(\frac{M_{Z^\prime}}{M_Z}\right)^2 
{\rm GeV} \; ,
\label{eq:zcouplimit}
\end{eqnarray}
\smallskip

Finally the cross section for the process
(\ref{eq:proc}) in this framework can be expressed as
\begin{equation}
\sigma_{\rm tot}= \sigma_{\rm SM}\, +\, 
G 
\, \sigma_{\rm int}(M_{Z^\prime},\Gamma_{Z^\prime})
\,+\,G^2 \,
\sigma_{Z^\prime}(M_{Z^\prime},\Gamma_{Z^\prime}) 
\label{eq:sigmatot}
\end{equation}
where the Standard Model, interference and new resonance contributions
are labeled SM, int and $Z^\prime$ respectively.  
\smallskip
 
\smallskip
%%%%%%%%%%%%%%%%%%%%%%%%%%%%%%%%%%%%%%%%%%%%%%%%%%%%%%%%%%%%%%%%%%%%%%%
%\section{Analyses details}
%\label{ana:frame}

The update of the bounds is based on the experimental analyses from
ATLAS~\cite{ATLASww_up} and CMS~\cite{CMSww_up}. There they analyzed the $W^+W^-$
production through the final state given in Eq.~\eqref{eq:proc}.
In our analyses we use the SM backgrounds that have been carefully evaluated by the
experimental collaborations and we only simulate the $Z^\prime$
signal and its interference with the SM. Nevertheless we also simulate the SM
production of $W^+W^-$ pairs in order to use this process to tune and validate our
Monte Carlo.\smallskip

In the original analysis~\cite{Eboli:2011ye} two different simulators were used and it was checked that
they led to compatible results with the experimental expectations for
the SM $W^+W^-$ pair production after tuning the simulators. We also checked
that both methods gave consistent results in the production of $Z^\prime$
signals and interferences for different points of the parameter space.
In order to update the bounds we use here what we labeled ``OUR ME-MC'', that is based
on the scattering amplitudes for the relevant processes obtained from the
package MADGRAPH~\cite{madevent}, while the evaluation was made with a {\sl
  homemade} Monte Carlo that evaluates the process \eqref{eq:proc} at the
parton level using the ${\cal O}(\alpha^4)$ signal matrix elements for
the subprocesses $q \bar{q} \to \ell^+ \nu \ell^{\prime -}
\nu^\prime$, with $\ell/\ell^\prime = e,\mu$. We used CTEQ6L parton
distribution functions \cite{CTEQ6} and the MADEVENT~\cite{madevent} default
renormalization and factorization scales.

\subsection{ATLAS analysis}

In order to account for some of the features included in the ATLAS evaluation of the SM $W^+W^-$
production, for instance highest available order simulations or detailed
detector simulations, we tune our simulator to obtain a total cross section for the different
flavor channels $ee$, $e\mu$ and $\mu\mu$ in the SM $W^+W^-$ process equal to the one in Table 5
of the ATLAS analysis~\cite{ATLASww_up} after the same cuts have been implemented.
The overall factors to tune our Monte Carlo are shown in Table~\ref{norma}.
The cuts in the ATLAS analysis are:

%%%%%%%%%%%%%%%
\begin{table}
\begin{tabular}{|c|c|c|c|c|}
\hline
Experiment & Monte Carlo & $ee$   & $e\mu$   & $\mu\mu$ 
\\
\hline
ATLAS  & OUR ME-MC & 0.51 & 0.70 & 0.92
\\
\hline
CMS  & OUR ME-MC & 0.56 & 0.83 & 0.95
\\
\hline
\end{tabular}
\caption{Overall multiplicative factors used to tune our simulator
  to the total number of events in the different flavour channels 
  predicted by the ATLAS and CMS simulations.}
\label{norma}
\end{table}

%%%%%%%%%%%%%%%

\begin{eqnarray}
&& |\eta_e|<1.37 \hbox{\ or\ }
1.52<|\eta_e|<2.47\hbox{\ and\ }|\eta_\mu|<2.4 .  
\label{cuts:atlas1}
\end{eqnarray}
\begin{eqnarray}
&& \Delta R_{ee}>0.3\hbox{\ and\ }\Delta R_{e\mu,\mu\mu}>0.2 \;.
\label{cuts:atlas2}
\end{eqnarray}
The transverse momentum cuts are slightly changed with respect to the original reference.
For the update events are selected if the leading lepton in each channel and the electron in the $e\mu$
channel accomplish
\begin{eqnarray}
&& p_T > 25 \hbox{ GeV,}
\label{cuts:atlas3}
\end{eqnarray}
while for the rest of leptons
\begin{eqnarray}
&& p_T > 20 \hbox{ GeV.}
\label{cuts:atlas4}
\end{eqnarray}
The cuts on the relative missing energy have also been increased with respect to the original analysis:
\begin{eqnarray}
&& M_{ee,\mu\mu} > 15 \hbox{ GeV } \;\;,\;\; M_{e\mu} > 10 \hbox{ GeV},
\nonumber
\\
&& | M_{ee,\mu\mu} -M_Z| > 15 \hbox{ GeV, } 
\label{cuts:atlas5}
\\
&& E_{T,~rel}^{miss}(ee)>50\hbox{ GeV } \;\;,\;\;E_{T,~rel}^{miss}(\mu\mu)>55\hbox{ GeV }
\nonumber\\
&& \hbox{and   }
E_{T,~rel}^{miss}(e\mu)>25\hbox{ GeV }, 
\nonumber
\end{eqnarray}
where $M_{\ell\ell}$ is the invariant mass of the lepton pair
and the relative missing energy is defined as:
\begin{equation}
E_{T,~rel}^{miss} =\left\{\begin{array}{clc}
&E_T^{miss}\times\sin{\Delta\phi_{\ell,j}}\ &\hbox{if}\ \ \Delta\phi_{\ell,j}<\pi/2\\
		     &E_T^{miss}\ &\hbox{if}\ \ \Delta\phi_{\ell,j}>\pi/2
                     \end{array} \right.
\label{etmissrel}
\end{equation}
with $\Delta\phi_{\ell,j}$ being the difference in the azimuthal angle
$\phi$ between the transverse missing energy and the nearest lepton or
jet. In a more detailed analysis jets would still have to be directly vetoed if $p_T>30$ GeV and
$|\eta_j|<4.5$.\smallskip

%%%%%%%%%%%%%%%
\begin{figure}[ht!]
\includegraphics[width=0.49\textwidth]{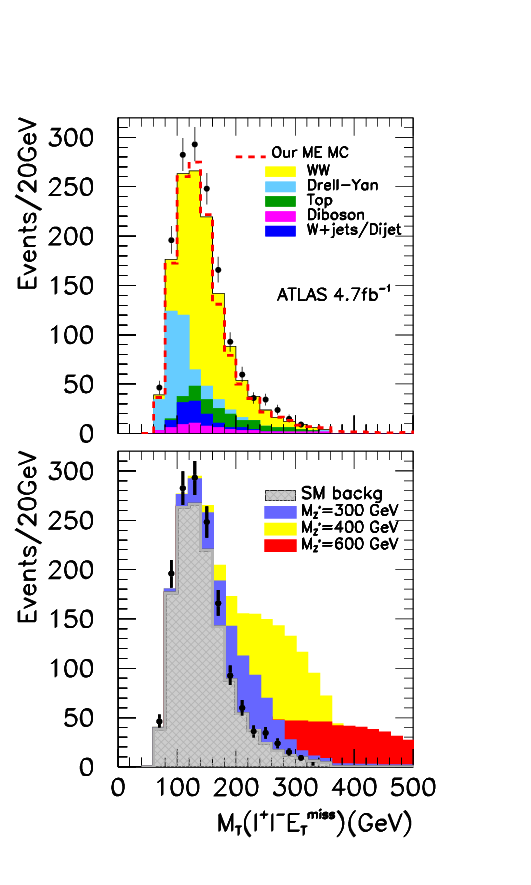}
\includegraphics[width=0.5\textwidth]{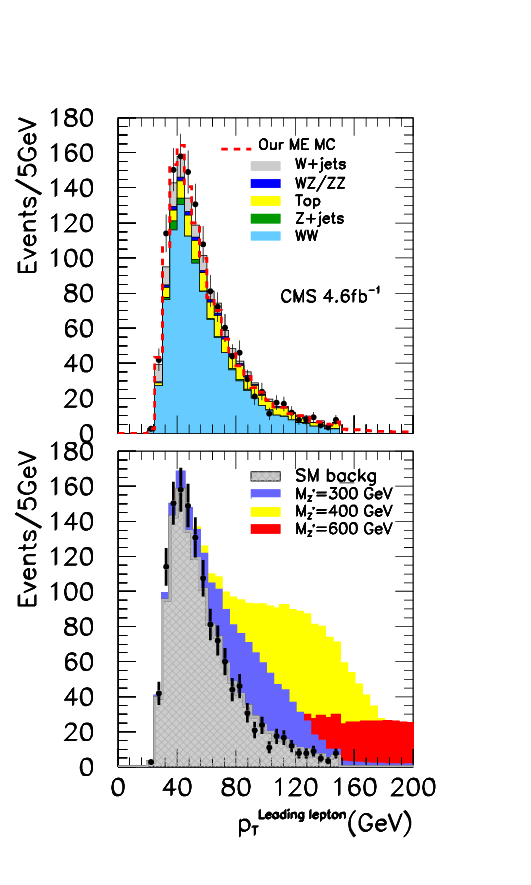}
\caption{Left upper panel: Transverse mass distribution of the SM
  contributions to the process $pp \to \ell^+ \ell^{\prime
    -}\sla{E}_T$ calculated by ATLAS (colored histograms) together
  with the number of observed events by ATLAS (points with error bars)
  and the performance of OUR ME-MC (red dashed). The results shown correspond to an integrated luminosity
  of $\mathcal{L}=4.7\ \mbox{fb}^{-1}$.
  \\
  Left lower panel: Transverse mass distribution of the total SM
  contribution to the process $pp \to \ell^+ \ell^{\prime -}\sla{E}_T$
  (gray hashed) together with the total expected number of events
  including a $Z^\prime$ of 300 GeV with $G=0.5$ (blue), a $Z^\prime$
  of 400 GeV with $G=1$ (yellow) and a $Z^\prime$ of 600 GeV with $G=1$
  (red).  For the three masses $\Gamma_{Z^\prime}=0.06M_{Z^\prime}$.
  We include also the ATLAS observed spectrum. Integrated luminosity
  of $\mathcal{L}=4.7\ \mbox{fb}^{-1}$.
  \\
  Right upper panel: Leading lepton transverse momentum distribution of the SM
  contributions to the process $pp \to \ell^+ \ell^{\prime
    -}\sla{E}_T$ calculated by CMS (colored histograms) together with
  the number of observed events by CMS (points with error bars) 
  and the performance of OUR ME-MC (red dashed). 
The results shown correspond to an integrated
luminosity of $\mathcal{L}=4.6\ \mbox{fb}^{-1}$.
\\
  Right lower panel: Transverse momentum of the leading lepton for the
  total SM contribution to the process $pp \to \ell^+ \ell^{\prime
    -}\sla{E}_T$ (gray hashed) together with the total expected number
  of events   including a $Z^\prime$ of 300 GeV with $G=0.5$  (blue), a
  $Z^\prime$ of 400 GeV with 
  $G=1$ (yellow) and a $Z^\prime$ of 600 GeV with $G=1$ (red). 
For the three masses  $\Gamma_{Z^\prime}=0.06M_{Z^\prime}$. 
We include also the observed distribution of events in CMS.
Integrated luminosity of $\mathcal{L}=4.6\ \mbox{fb}^{-1}$.}
\label{fig:dist_atlas}
\end{figure}
%%%%%%%%%%%%%%%

After our Monte Carlo has been tuned, we compare the transverse mass
of the SM $W^+W^-$ pair
production from our simulator with the expectation from ATLAS in order to validate our Monte Carlo.
Both distributions can be found in the left upper panel of
Fig.~\ref{fig:dist_atlas}. The results shown correspond to an
integrated luminosity of $\mathcal{L}=4.7\ \mbox{fb}^{-1}$. In the figure the stacked histograms
for the different background processes as expected by ATLAS collaboration are shown together with
our SM $W^+W^-$ production expectation added to the ATLAS results for the rest of backgrounds (red dashed).
It can be seen that our simulation approximates very well the ATLAS expectation.
\smallskip

As in the original analysis we use the normalization factors obtained
from SM $W^+W^-$ pair production to simulate the $Z^\prime$ signal and interference.
As an illustration of the effects of including a new neutral vector resonance
in the transverse mass spectrum of the process we show the expected $M_T$ distribution
for three different $Z^\prime$ new resonances after applying all the cuts and for a integrated luminosity
of 4.7 fb$^{-1}$ in the left lower panel of Figure~\ref{fig:dist_atlas}.
It can be seen that the effect of new spin--1 neutral resonances is
characterized by an excess of events with respect to the SM expectations at the higher values
of $M_T$.
\smallskip

Given this behavior we use the transverse mass spectrum to place
constraints on the $Z^\prime$ properties. We construct a binned log-likelihood
function based on the contents of the different bins in the $M_T$ distribution and
we assume the number of observed events follow independent Poisson
distributions in each bin. The details of the statistical analysis can be found
in the published reference~\cite{Eboli:2011ye}. The pulls~\cite{pulls} used to estimate
the effect of systematic uncertainties are updated from Table 5 of~\cite{ATLASww_up} to:
\begin{eqnarray}
\sigma^{st}_b&=&0.026\ \ \ \ \ \ \sigma^{sy}_b=0.09\\
\sigma^{st}_s&=&0.005\ \ \ \ \ \ \sigma^{sy}_s=0.10
\label{eq:pulls_atlas}
\end{eqnarray}

The only change in the analysis with respect to the original publication~\cite{Eboli:2011ye}
is that the upper limit of the ATLAS transverse mass distribution has been increased from 340 GeV to 360 GeV.
We then performed two analyses:
in the first one we computed the likelihood with the 16 transverse mass bins in~\cite{ATLASww_up}
between $M_T=40$ GeV and $M_T=360$ GeV (\ie\ $N^{max}_{AT}=16$), while in
the second one we added an extra 17th bin (\ie\ $N^{max}_{AT}=17$) where
we assumed that the number of observed events and SM expected
predictions are null and where we added the $Z^\prime$ expected contributions with $M_T>360$ GeV.
\smallskip

\subsection{CMS analysis}

In the case of the CMS analysis the details of the simulation are analogous to the ATLAS ones.
We tune our homemade Monte Carlo to account for the different details of the simulation
by comparing the SM $W^+W^-$ pair production
in the $ee$, $e\mu$, and $\mu\mu$ channels with respect to the expectations presented in
Ref.~\cite{CMSww_up}. The cuts in the new CMS reference are:
\begin{eqnarray}
 &&|\eta_e|<2.5 \hbox{ and }|\eta_\mu|<2.4,
\label{cuts:cms1}
\\
&&\Delta R_{ee}>0.4 \hbox{ and }\Delta R_{e\mu,\mu\mu}>0.3.
\label{cuts:cms2}
\end{eqnarray}
The leptons need to verify also:
\begin{eqnarray}
&&p_T^{\rm leading} > 20 \hbox{ GeV,}
\nonumber
\\
&&p_{T\ e\mu}^{\rm subleading} > 10 \hbox{ GeV,   } p_{T\ ee,\mu\mu}^{\rm subleading} > 15 \hbox{ GeV}
\nonumber
\\
&&M_{ee,\mu\mu} > 20 \hbox{ GeV }\;\;\hbox{and}\;\; M_{e\mu} > 12 \hbox{ GeV,}
\label{cuts:cms3}
\\
&&| M_{ee,\mu\mu} -M_Z| > 15 \hbox{ GeV, }
\nonumber
\\
&&E_{T,~rel}^{miss}(ee,\mu\mu)>40\hbox{ GeV } \;\;\hbox{and}\;\;E_{T,~rel}^{miss}(e\mu)>20\hbox{ GeV.}
\nonumber
\end{eqnarray}
Comparing the new and old CMS analyses, one can notice
that the requirement on the transverse momentum of the subleading lepton has been increased in order to
reduce the low-mass $Z/\gamma^*\rightarrow \ell^+\ell^-$ contribution and the $W+$jets background in the $ee$
and $\mu\mu$ channels. Furthermore, the cut on the minimum $M_{ee,\mu\mu}$ is also stronger
in order to suppress contributions from low mass resonances. Finally a new cut in the transverse
momentum of the system formed by the pair of leptons has been included in the
new CMS reference for all three channels $ee$, $e\mu$ and $\mu\mu$:
\begin{eqnarray}
&&  p_T^{\ell\ell}>45 \hbox{ GeV.}
\label{cuts:cms3y}
\end{eqnarray}
The aim of this new cut is to further reduce the contribution of Drell--Yan and fake background
contamination. It is worth noting that the jet veto and the cut on the angle in the transverse
plane between the dilepton system and the most
energetic jet with $p_T>15$ GeV would still have to be directly applied when doing a more detailed simulation.
\smallskip

As in the ATLAS case the normalization factors needed to tune our simulator after applying all the cuts
are shown in Table~\ref{norma}. In order to validate our simulator we compare the SM background expectations
from CMS collaboration and the sum of our SM $W^+W^-$ pair production simulation to the rest of CMS backgrounds
(red dashed) as a function of the transverse momentum of the leading lepton
in the right upper panel of Fig.~\ref{fig:dist_atlas}.
Our simulation approximates very well the CMS expectations.
The effect of introducing new neutral vector resonances can be observed in the right lower panel of
the same Fig.~\ref{fig:dist_atlas}. For the simulation of the $Z^\prime$ signal and the interference we employed the
same normalization factors obtained from the $W^+W^-$ SM production
for the channels $ee$, $e\mu$, and $\mu\mu$. The presence of the new
$Z^\prime$ enhances the contribution at the higher values of the
transverse momentum of the leading lepton. Consequently the exclusion
limits on the production of a $Z^\prime$ were extracted
using a binned log-likelihood function based on the contents of the
bins of the transverse momentum distribution of the leading
lepton~\cite{Eboli:2011ye}.
\smallskip

As in the original reference we performed two analysis. First we calculated
the binned log--likelihood function using the events in the bins shown in
the CMS image. That means the event rates in the 26 leading lepton transverse
momentum bins between 20 GeV and 150 GeV (\ie\ $N^{max}_{CMS}=26$).
In the second analysis we added an extra bin where we included the number of observed
events and background expectations that are left with values higher than 150 GeV.
These values can be obtained from comparing
the quantities read from the images with the values quoted in Table 2 of~\cite{CMSww_up}. In this extra bin
we also added the expected
contributions from the $Z^\prime$ with $p_T^{\rm leading}>150$ GeV (\ie\ $N^{max}_{CMS}=27$).

\subsection{Combined Analysis}

In order to get more stringent bounds on the production of a $Z^\prime$
that decays into electroweak gauge boson pairs
we combined the ATLAS and CMS results by constructing the
combined log-likelihood function assuming conservatively
that the ATLAS and CMS systematic uncertainties are uncorrelated.\smallskip

In all cases  we set the 2$\sigma$ exclusion limits (2$\sigma$, 1 d.o.f)
on $G$ by maximizing the corresponding likelihood function (or 
equivalently minimizing the $\chi^2$) with respect to $G$ 
for each value of $M_{Z'}$ and $\Gamma_{Z'}$ 
and imposing
\begin{eqnarray}
|\chi^2 (M_{Z'},G,\Gamma_{Z'})-\chi^2_{\rm min}(M_{Z'},\Gamma_{Z'})|>4 \; .
\end{eqnarray}

%%%%%%%%%%%%%%%%%%%%%%%%%%%%%%%%%%%%%%%%%%%%%%%%%%%%%%%%%%%%%%%%%%%%%%
\section{Results}
\label{sec:results}

All the $2\sigma$ exclusion limits on new neutral vector resonances
that decay to electroweak gauge boson pairs are shown in the plane
$G \otimes M_{Z^\prime}$ for three possible values of the $Z^\prime$
width $\Gamma_{Z'}/M_{Z'}=0.01$, $0.06$ and $0.3$.\smallskip

%--------------
\begin{figure*}
\includegraphics[width=0.7\textwidth]{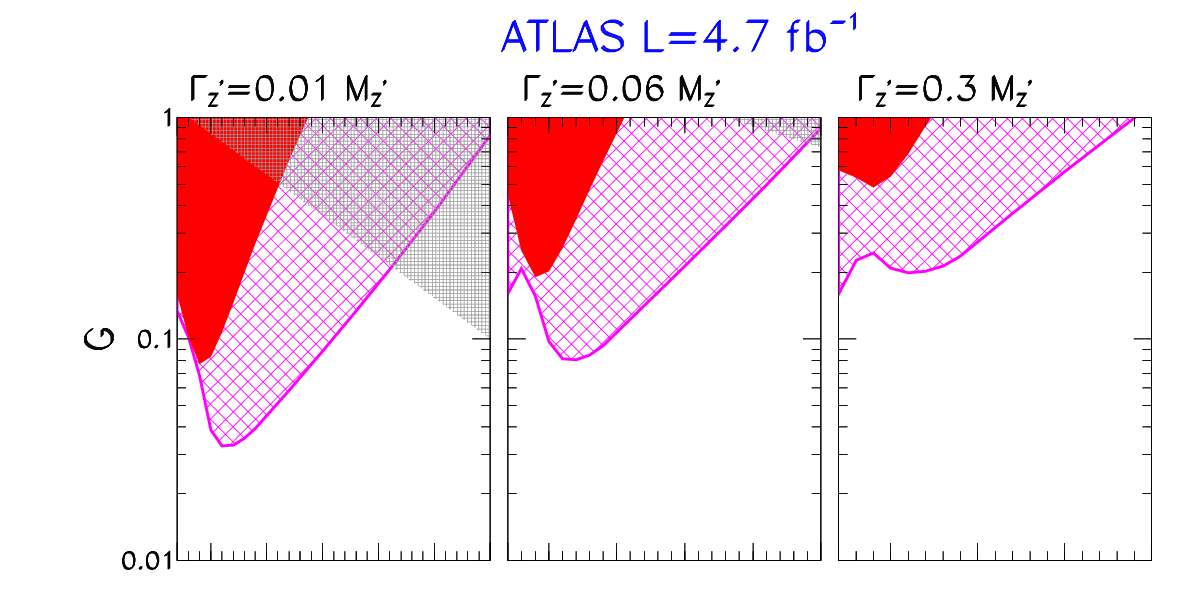}
\vglue -0.65cm
\includegraphics[width=0.7\textwidth,angle=180]{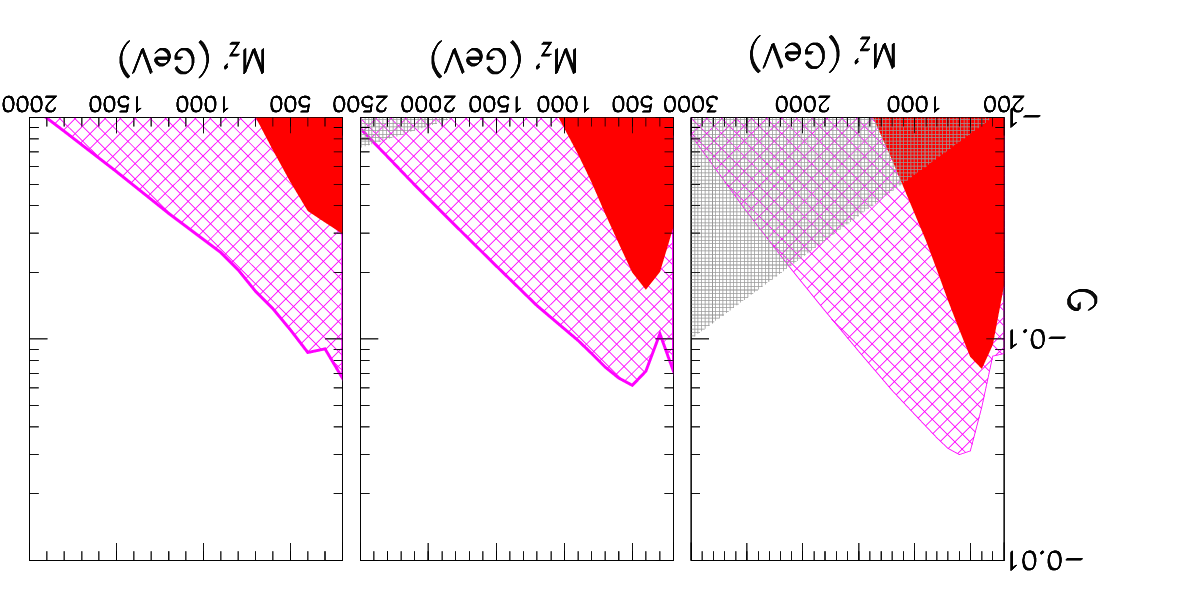}
\caption{ 2$\sigma$ exclusion limits on the production of a $Z^\prime$
  from our analysis of the $M_T$ distribution measured by ATLAS with
  $\mathcal{L}=4.7\ \mbox{fb}^{-1}$ and for three values of
  $\Gamma_{Z'}/M_{Z'}=0.01$, $0.06$ and $0.3$ (left, center and right
  panels respectively).  The red solid regions are derived using the
  log-likelihood function with
  $N^{max}_{AT}=16$. The purple hatched regions are derived using the
  log-likelihood function with
  $N^{max}_{AT}=17$.  The shadowed regions in the upper (lower) right
  corner of the upper (lower) panels represent the excluded values by
  the condition Eq.~\eqref{eq:zcouplimit}.}
\label{fig:bounds_atlas}
\end{figure*}
%-------------

The bounds for the ATLAS analysis, corresponding to the study of
the transverse mass spectrum observed with an integrated luminosity
of  $\mathcal{L}=4.7\ \mbox{fb}^{-1}$ are presented in Fig.~\ref{fig:bounds_atlas}.
There we can distinguish three different regions: the
grey shadowed regions in the upper right (lower right) of the upper (lower) panel
correspond to points excluded by requiring the consistency of the total decay width
of a $Z^\prime$ with its decay to light quarks and SM $W^+W^-$ pairs as expressed in Eq.~\eqref{eq:zcouplimit}.
The red solid regions were derived making the analysis with $N^{max}_{AT}=16$ bins,
between $M_T=40$ GeV and $M_T=360$ GeV, while the purple hatched regions
contain the points excluded  when the extra bin accounting for values
of transverse mass $M_T>360$ GeV is included, $N^{max}_{AT}=17$.
One can observe that the bounds
are stronger for narrow resonances, while including the extra bin has a bigger impact
for a wider and heavier $Z^\prime$. Furthermore the effect of the interference, that can be observed by comparing
the upper and lower panels, is noticeable only for wider and lighter new resonances, as expected
from the interference term being roughly proportional to
$\Gamma_{Z^\prime}/M_{Z^\prime}$.  
\smallskip

%--------------
\begin{figure*}
\includegraphics[width=0.7\textwidth]{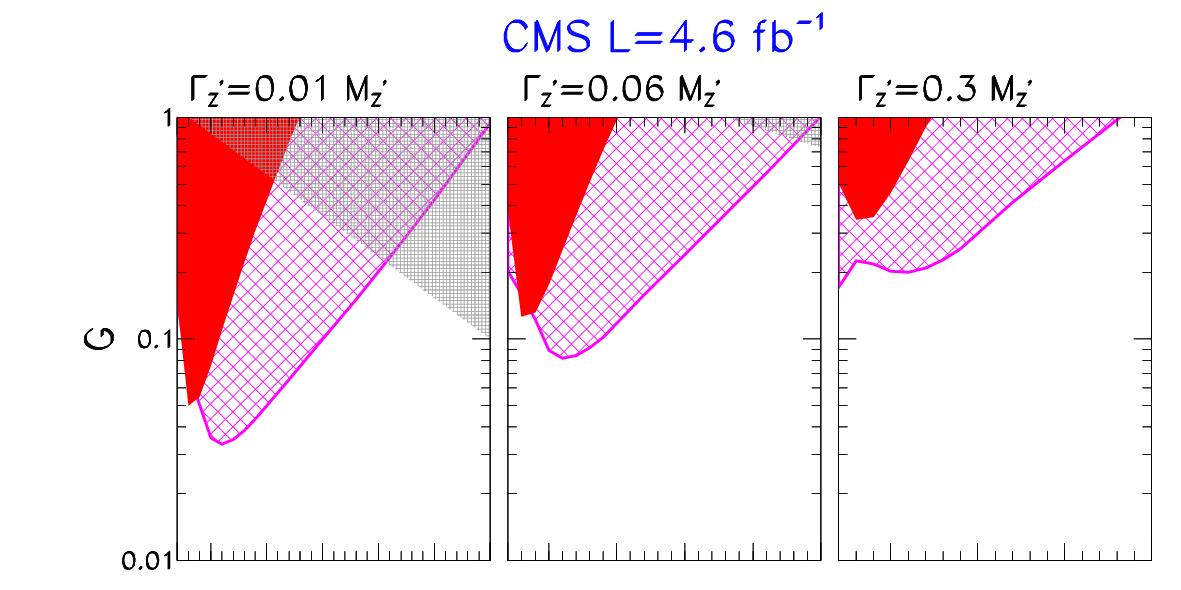}
\vglue -0.6cm
\includegraphics[width=0.7\textwidth,angle=180]{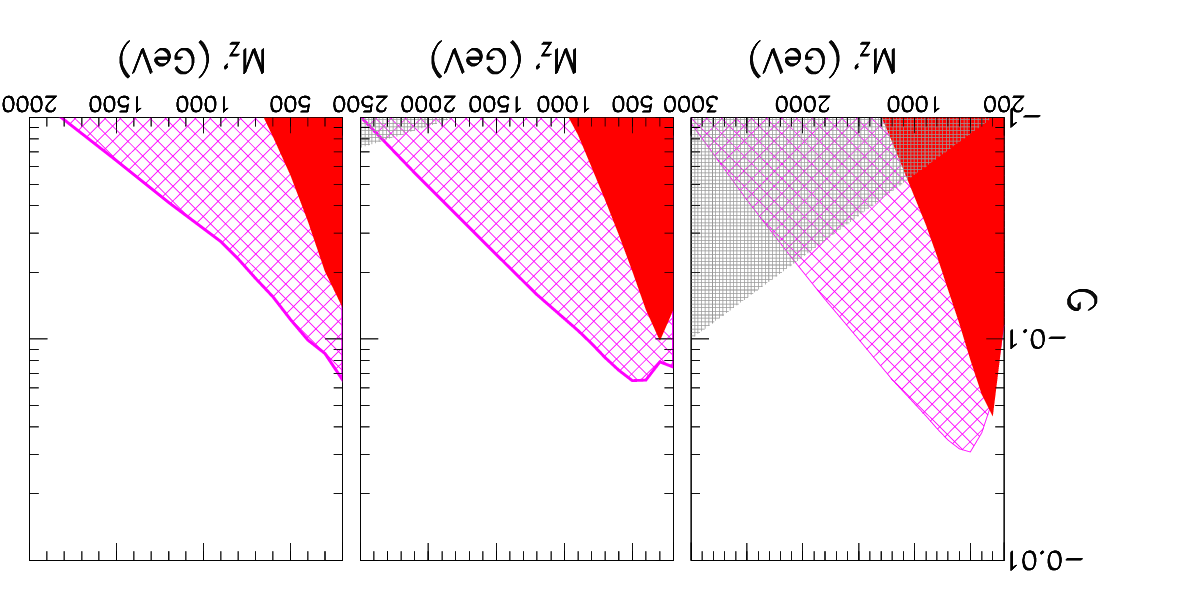}
\caption{ 2$\sigma$ exclusion limits on the production of a $Z^\prime$
  from our analysis of the $p_T^{\rm leading}$ distribution measured
  by CMS with $\mathcal{L}=4.6\ \mbox{fb}^{-1}$.  The left, center
  and right panels correspond to three values of
  $\Gamma_{Z'}/M_{Z'}=0.01$ ,$0.06$ and $0.3$ respectively.  The red
  solid regions are derived using the log-likelihood function
  with $N^{max}_{CMS}=26$.  The purple hatched
  regions are derived using the log-likelihood function
  with $N^{max}_{CMS}=27$.  The shadowed
  regions in the upper (lower) right corner of the upper (lower)
  panels represent the excluded values by the condition
  Eq.~\eqref{eq:zcouplimit}.}
\label{fig:bounds_cms}
\end{figure*}
%--------------

In the CMS case the $2\sigma$ exclusion limits on the production of a $Z^\prime$
derived from the analysis of the $p_T^{\rm leading}$ distribution
measured with an integrated luminosity of $\mathcal{L}=4.6\ \mbox{fb}^{-1}$ can be seen in
Fig.~\ref{fig:bounds_cms}. The bounds are very similar to the ATLAS case. The only difference
is in the shape of the exclusion limits without the extra bin. This is
due to the fact that within the range of the kinematic values and the kinematic variables used
CMS is more sensitive than ATLAS to the lightest masses when no extra bin is included.
\smallskip

%--------------
\begin{figure*}
\includegraphics[width=0.7\textwidth]{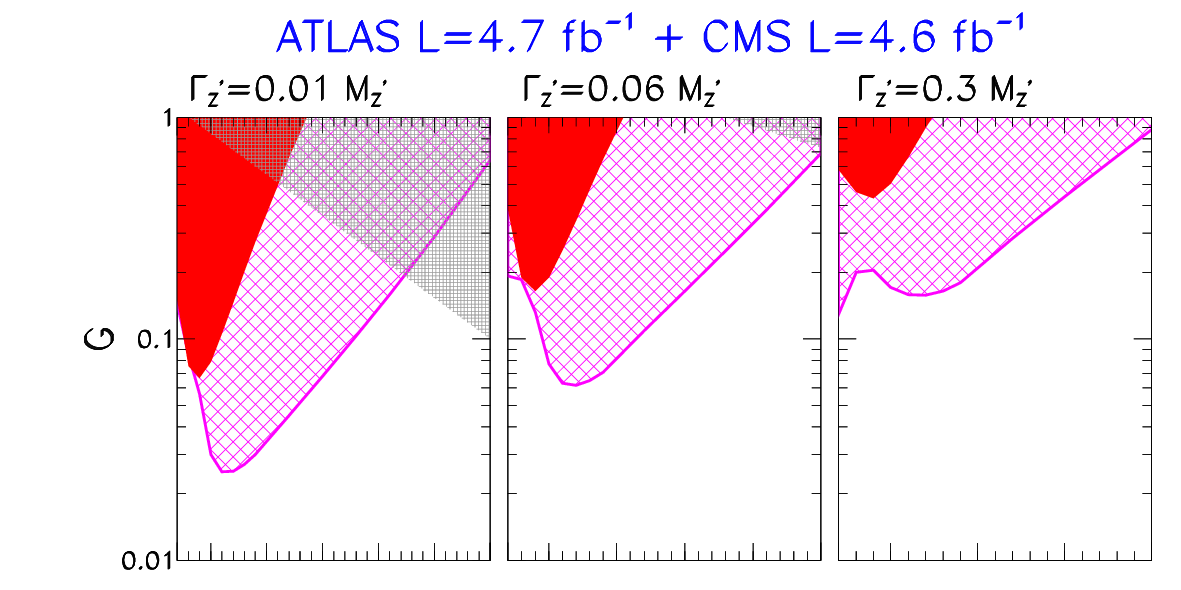}
\vglue -0.6cm
\includegraphics[width=0.7\textwidth,angle=180]{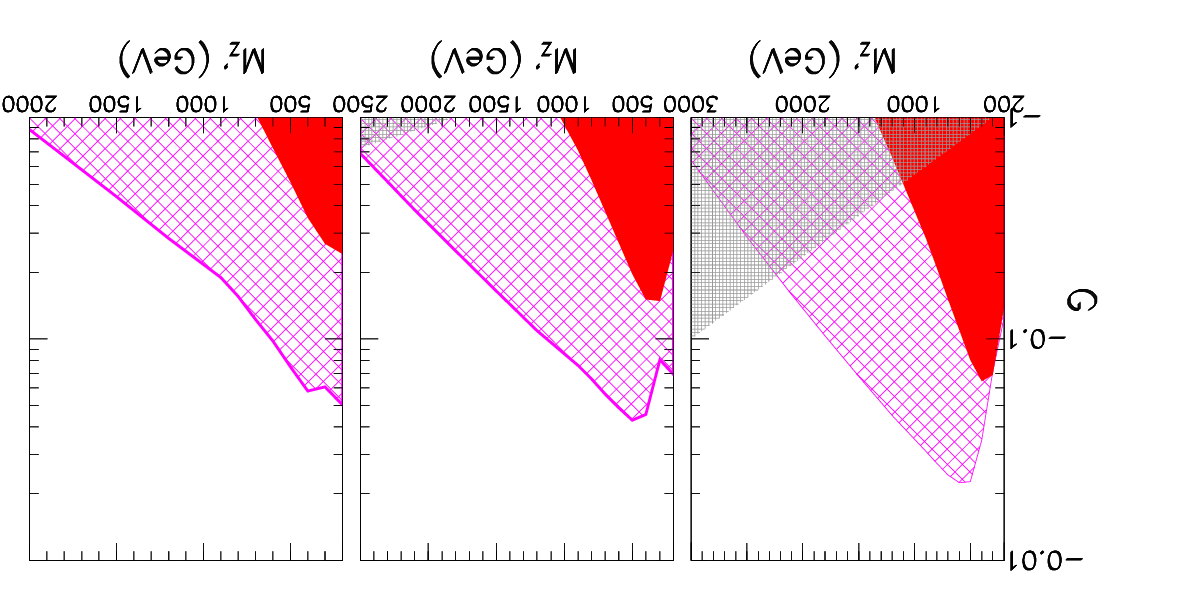}
\caption{2$\sigma$ exclusion limits on the production of a $Z^\prime$
  from our combined analysis of the measured $M_T$ distribution in
  ATLAS with $\mathcal{L}=4.7\ \mbox{fb}^{-1}$ and the $p_T^{\rm
    leading}$ distribution measured by CMS with $\mathcal{L}=4.6\
  \mbox{fb}^{-1}$.  The red solid (purple hatched) regions are derived
  using the combined log-likelihood function
  with 16 and 27 (17 and 27) bins of the ATLAS and CMS distributions
  respectively.  The shadowed regions in the upper (lower) right
  corner of the upper (lower) panels represent the excluded values by
  the condition Eq.~\eqref{eq:zcouplimit}.}
\label{fig:bounds_comb}
\end{figure*}
%--------------

Finally the $2\sigma$ exclusion limits on the production
of new $Z^\prime$ from the combination of the analysis of the
transverse mass spectrum in ATLAS with an integrated luminosity
of $\mathcal{L}=4.7\ \mbox{fb}^{-1}$ and the $p_T^{\rm leading}$
distribution spectrum in CMS with $\mathcal{L}=4.6\ \mbox{fb}^{-1}$ are shown
in Fig.~\ref{fig:bounds_comb}.
These are the strongest existing direct bounds on the production
of new neutral vector resonances that decay to electroweak gauge boson pairs.
Comparing to the original reference~\cite{Eboli:2011ye}
where $\mathcal{L}=1.02\ \mbox{fb}^{-1}$ from ATLAS and $\mathcal{L}=1.55\ \mbox{fb}^{-1}$
from CMS were used, we can observe that now
from our combined analysis with 17 and 27 bins from the ATLAS and CMS
distributions respectively, a narrow resonance of any mass with
$\Gamma_{Z'}/M_{Z'}=0.01$ and that saturates the partial wave
amplitude for the process $W^+ W^- \to W^+ W^-$ is excluded at 2$\sigma$ level
if its coupling to the light quarks is larger than 19\% of
the SM $Z\bar{q}q$ coupling. From the extended analysis we can also see that
a new neutral vector resonance that saturates the
partial wave amplitude for the process $W^+ W^- \to W^+ W^-$ and
couples to light quarks with SM strength is completely excluded for
$\Gamma_{Z'}/M_{Z'}=0.01$ and $\Gamma_{Z'}/M_{Z'}=0.06$, while for
a wider resonance, $\Gamma_{Z'}/M_{Z'}=0.3$, it is
excluded for masses up to 2 TeV.
\smallskip

As it is shown in the original analysis~\cite{Eboli:2011ye}, the bounds there are already generically stronger
than the ones obtained by the CDF collaboration analyzing $WW$ production
at the Tevatron~\cite{cdf}. Looking at the exclusion limits figures presented here we can observe
that with the new $\mathcal{L}=4.7\ \mbox{fb}^{-1}$ and $\mathcal{L}=4.6\ \mbox{fb}^{-1}$
data sets from ATLAS and CMS respectively, the exclusion limits not only are extended
to heavier masses but also for a given mass and
width the couplings excluded are extended to values around 60$\%$ of the original
exclusion limits.
\smallskip

All the $2\sigma$ exclusion limits on the production of a new $Z^\prime$
are presented in a model independent way. Consequently they can be translated to many
different EWSB models. A detailed example of how to translate the bounds to a given model
can be found in the published analysis~\cite{Eboli:2011ye}.\smallskip
\smallskip

%%%%%%%%%%%%%%%%%%%%%%%%%%%%%%%%%%%%%%%%%%%%%%%%%%%%%%%%%%%%%%%%%%%%%%%
%\section{Summary}
%\label{sec:summary}

To summarize, in this talk the update of the bounds on new neutral vector resonances in electroweak
gauge boson pair production using $\mathcal{L}=4.7\ \mbox{fb}^{-1}$ from ATLAS
and $\mathcal{L}=4.6\ \mbox{fb}^{-1}$ from CMS is presented. The exclusion limits
on the production of a $Z^\prime$ associated with the EWSB sector are placed
analyzing the kinematic distributions of the $pp\to\ell^+\ell^{\prime -}\sla{E}_T$ events.
The model independent results correspond to the most
stringent exclusion limits available on the production of
a $Z^\prime$ that decay to SM $W^+W^-$ pairs,
well exceeding the
limits from Tevatron and the previous LHC limits.
\smallskip
\smallskip

J.G-F is supported by MICINN
FPA2010-20807, consolider-ingenio 2010 program CSD-2008-0037,
EU grant FP7 ITN INVISIBLES (Marie Curie Actions PITN-GA-2011-289442)
and by Spanish ME FPU grant AP2009-2546.
%%%%%%%%%%%%%%%%%%%%%%%%%%%%%%%%%%%%%%%%%%%%%%%%%%%%%%%%%%%%%%%%%%%%%%

\bibliographystyle{h-physrev4}

\end{document}